\documentclass[11pt,oneside,letterpaper]{article}
\usepackage{amssymb}
\usepackage{amsmath}
\usepackage[dvips]{graphicx}
\usepackage{setspace}
\usepackage{amsfonts}
\usepackage{fancyhdr}
\usepackage{xcolor}
\usepackage{graphicx}
\usepackage{rotating}
\usepackage{comment}
\usepackage{color}
\usepackage{cite}
\usepackage{braket}
\definecolor{darkgreen}{rgb}{0,0.5,0}
\definecolor{darkblue}{rgb}{0,0,0.6}
\definecolor{purple}{rgb}{0.4,.2,0.7}
\newcommand{\p}{\partial}

\newcommand{\f}{\frac}

\newcommand{\be}{\begin{equation}}
\newcommand{\ee}{\end{equation}}

\newcommand{\tbar}{\widetilde{T}}
\usepackage[colorlinks=true,citecolor=darkgreen,linkcolor=black,urlcolor=purple]{hyperref}

\usepackage{pdfsync}

\makeatletter
\newcommand*{\defeq}{\mathrel{\rlap{%
                     \raisebox{0.3ex}{$\m@th\cdot$}}%
                     \raisebox{-0.3ex}{$\m@th\cdot$}}%
                     =} 
\makeatother

\DeclareMathOperator{\e}{\epsilon}

\def\be{\begin{eqnarray}}
\def\ee{\end{eqnarray}}

\newcommand{\bea}{\begin{eqnarray}}
\newcommand{\eea}{\end{eqnarray}}
\def\ben{\begin{equation}}
\def\een{\end{equation}}
\def\half{{\textstyle{\frac{1}{2}}}}

\let\a=\alpha  \let\g=\gamma \let\d=\delta \let\e=\varepsilon
   
\let\l=\lambda    \let\p=\phi \let\r=v
 \let\t=\tau

 \let\G=\Gamma \let\D=\Delta

\let\f=\frac

\let\pa=\partial
\def\be{\begin{equation}}
\def\ee{\end{equation}}
\def\ba{\begin{array}}
\def\ea{\end{array}}
\def\del{\partial}

\def\ba#1\ea{\begin{align}#1\end{align}}
\def\bs#1\es{\begin{split}#1\end{split}}

\renewcommand{\p}{\partial}

\allowdisplaybreaks  

\numberwithin{equation}{section}

\begin{document}
\onehalfspacing

\begin{center}

~
\vskip5mm

{\LARGE  {
Holography at finite cutoff with a $T^2$ deformation
}}

\vskip10mm

Thomas Hartman,$^{1}$\ \ Jorrit Kruthoff,$^{2}$\ \  Edgar Shaghoulian,$^{1}$\ \  and Amirhossein Tajdini$^{1}$

\vskip5mm

{\it 1) Department of Physics, Cornell University, Ithaca, New York, USA} \\
\vskip5mm
{\it 2) Institute for Theoretical Physics Amsterdam and Delta Institute for Theoretical Physics, University of Amsterdam, Science Park 904, 1098 XH Amsterdam, The Netherlands} 

\vskip5mm

\tt{ hartman@cornell.edu, j.kruthoff@uva.nl, eshaghoulian@cornell.edu, at734@cornell.edu}

\end{center}

\vspace{4mm}

\begin{abstract}
\noindent
We generalize the $T\overline{T}$ deformation of CFT$_2$ to higher-dimensional large-$N$ CFTs, and show that in holographic theories, the resulting effective field theory matches semiclassical gravity in AdS with a finite radial cutoff. We also derive the deformation dual to arbitrary bulk matter theories. Generally, the deformations involve background fields as well as CFT operators. By keeping track of these background fields along the flow, we demonstrate how to match correlation functions on the two sides in some simple examples, as well as other observables.

 \end{abstract}

\pagebreak
\pagestyle{plain}

\setcounter{tocdepth}{2}
{}
\vfill
\tableofcontents

\newpage

\section{Introduction}
Quantum gravity in finite volume is a difficult problem that is perhaps vital to fundamental cosmology. A natural question is how to apply holographic duality in this context. The avenue we will explore is to impose a hard radial cutoff in AdS and approach this problem as a deformation of AdS/CFT.

The precise relation between a radial cutoff in the bulk geometry and a cutoff in the boundary field theory is a longstanding puzzle in AdS/CFT, discussed since the advent of the duality itself. The UV/IR relation \cite{Susskind:1998dq} of the duality  provides a clue but is far from a precise relationship. In addition to being an important entry in the AdS/CFT dictionary, finding such a relationship may prove fruitful in decoding local physics in the bulk and in constructing a framework for holography in more general spacetimes.

Most of the work on this topic has focused on understanding the long distance physics of the original CFT, in the spirit of the renormalization group \cite{deBoer2000,Bianchi:2001kw, Heemskerk:2010hk,Faulkner:2010jy, Balasubramanian:2012hb, Lee:2013dln}. Recently, a different perspective was emphasized in \cite{McGough:2016lol}, in the context of pure 3d gravity. Here the goal is not to understand the original CFT, but to explicitly deform the CFT so that it reproduces the bulk physics with Dirichlet boundary conditions at finite cutoff. This is a deformed theory in the bulk, dual to a deformed theory on the boundary. The proposal in \cite{McGough:2016lol} is that 3d gravity at finite radial cutoff is dual to a 2d CFT deformed by the irrelevant operator $T\bar{T}$, a deformation previously studied in the field theory context by Zamolodchikov and Smirnov \cite{Zamolodchikov:2004ce,SmirnovZamolodchikov}. The analytic tractability of CFT deformations by this operator, which follows primarily from the Zamolodchikov factorization equation, allows nontrivial checks of the proposal. This deformation and its holographic interpretation were explored further in \cite{Cavaglia:2016oda, Dubovsky:2017cnj, Shyam:2017znq, Kraus:2018xrn, Cardy:2018sdv, Aharony:2018vux, Dubovsky:2018bmo, Donnelly:2018bef, Bonelli:2018kik, Chen:2018eqk}.

In this paper, we propose an effective field theory (EFT) dual to a general bulk theory at finite cutoff, generalizing the $T\bar{T}$ deformation to higher dimensions and allowing for matter couplings. In the field theory, the tool that will replace the factorization property of $T\bar{T}$ is large-$N$ factorization.

We will first provide a recipe to derive the necessary CFT deformation for arbitrary bulk theories in AdS$_{d+1}$. Using this recipe, we will find the deformation in several examples. For example, for pure Einstein gravity in $d=3,4$, the deformation of the CFT is 
\be\label{introflow}
\f{\p S}{\p\lambda} =\int d^d x \sqrt{\gamma} \left((T_{ij}+b_d G_{ij})(T^{ij} + b_d G^{ij})-\f{1}{d-1}(T^{i}_{i}  + b_d G^i_i)^2 \right)\ ,
\ee
where $\lambda$ is a dimensionful coupling, $b_d \propto \lambda^{2/d-1}$ with a coefficient given below, and $G_{ij}$ is the Einstein tensor for the boundary metric. (The expression for arbitrary $d$ is below.)

This flow equation, in which the stress tensor $T_{ij}$ also depends on $\lambda$, determines the classical action of the boundary EFT. The background terms in \eqref{introflow} induce a redefinition of the operator $T_{ij}$ along the flow, and are necessary to match correlation functions, even on flat backgrounds. We also derive the explicit flow equation for CFTs with scalar operators or at finite $U(1)$ charge density. The scalar case leads to an effective field theory deformed by ${\cal O}^2 - (\p J)^2$, which provides a simple toy model for \eqref{introflow}. The $U(1)$ case allows us to compare to charged black holes in Einstein-Maxwell theory. 

With the CFT deformations in hand we compute various quantities in the deformed CFT and compare to bulk AdS quantities at finite cutoff, finding perfect agreement for $\lambda > 0$. In particular, we will match the two-point correlation functions in vacuum, as well as the energy spectrum and thermodynamics.

A finite Dirichlet cutoff in the bulk is a dramatic, and perhaps violent, deformation of the gravitational theory. Intuitively, this is because gravity with reflecting boundary conditions induces negative image masses on the other side of the wall, which screen the gravitational force. This raises the possibility that the theory violates causality, as discussed in \cite{Marolf:2012dr,Andrade:2015gja,McGough:2016lol}, or that the dual EFT cannot be UV-completed as an ordinary quantum field theory for positive $\lambda$  (see e.g. \cite{Cardy:2015xaa}). We will sidestep these issues by restricting the discussion to physics below the cutoff, where both sides of the duality appear to make sense, at least perturbatively. Some speculations on the the UV are mentioned in the discussion section.

It is also possible to consider the deformation with $\lambda < 0$. In this case, the asymptotic density of states is super-Hagedorn in the UV, giving a scaling  $\log \rho \propto E^\f{2(d-1)}{d}$. Intriguingly, this scaling agrees with the density of states of $p$-branes (with $p = d-1$) in the semiclassical approximation \cite{Duff:1987cs, Fubini:1972mf, Strumia:1975rd, Alvarez:1991qs}. The matching with Hagedorn scaling in $d=2$ is an important aspect of relating (a single-trace version of) this deformation to little string theory \cite{Giribet:2017imm,Giveon:2017myj, Giveon:2017nie, Asrat:2017tzd, Chakraborty:2018kpr}.

The derivation starts with the Hamilton-Jacobi equation in the bulk, and uses the techniques of holographic renormalization developed in \cite{deBoer2000,deHaro2001,Bianchi:2001kw,Martelli:2002sp}. However, instead of trying to relate \eqref{introflow} to an RG equation, we view it as the definition of a boundary EFT that can be studied on its own terms. This is the perspective taken in \cite{McGough:2016lol}, in contrast to the earlier work cited above. 
This approach leaves open the mysterious question emphasized in \cite{Heemskerk:2010hk,Mintun:2014gua} of what coarse-graining or cutoff procedure in the QFT actually produces the flow \eqref{introflow}. If this procedure were known, then \eqref{introflow} would need to emerge from it automatically, whereas in our approach the bulk Hamiltonian must be input by hand.

Although the operator \eqref{introflow} has been derived from a bulk calculation, it is an EFT operator, giving a purely field-theoretic definition of the deformation, as in the 2d case. Like in 2d, the deformation is defined order by order in conformal perturbation theory in $\l$, but unlike in 2d, it is only unambiguously defined in perturbation theory in $1/N$. For $d=2$ a nonperturbative definition for the theory on Minkowski spacetime is provided by the S-matrix \cite{Dubovsky:2017cnj}, or at $c=24$ with certain sign of the coupling by critical string theory \cite{Caselle:2013dra}.

As this work was being completed, \cite{Taylor:2018xcy} appeared, which also derives the source-free versions of equation \eqref{introflow} and the corresponding equation \eqref{deformMatter} with $U(1)$ charge.

\subsection{The dictionary at finite cutoff}\label{ss:dictionary}

In the rest of this introduction, we will present our proposed dictionary for the EFT dual to a sharp radial cutoff in AdS. Begin by choosing coordinates
\be\label{icoord}
ds^2 = g_{\mu\nu}dx^\mu dx^\nu = N(r)^2 dr^2 + r^2 \gamma_{ij}dx^i dx^j \ ,
\ee
with $N(r) \to 1/r$ near the boundary and where we have set the AdS radius $\ell_{AdS}=1$. 
The usual AdS/CFT dictionary states
\be\label{regd}
Z_{CFT}[\gamma_{ij}, J] = 
\lim_{r_c \to \infty}
Z_{grav}[ g^0_{ij} = r_c^2 \gamma_{ij},
\phi_0 = r_c^{\Delta  - d} J ] \ .
\ee
On the left is the CFT generating function, in the metric $\gamma_{ij}$, with source $J$ for a scalar operator ${\cal O}$ of dimension $\Delta$. (Later we will generalize to spinning sources.) On the right is the gravitational path integral with the Dirichlet boundary conditions
\be
g_{ij}(r_c, x) = g^0_{ij}(x) , \quad
\phi(r_c, x) = \phi_0(x) \ .
\ee
In \eqref{regd}, we have inserted the explicit factors of the radial cutoff $r_c$ to ensure that CFT correlators, computed by $\left(\f{1}{\sqrt{\gamma}}\frac{\delta}{\delta J}\right)\left(\f{1}{\sqrt{\gamma}} \frac{\delta}{\delta \gamma_{ij}}\right) \cdots \log Z$, are normalized to be independent of $r_c$.

There is considerable arbitrariness in how the dictionary \eqref{regd} should be extended to finite $r_c$. We choose the simplest prescription, which is to assume that the same dictionary defines an effective boundary theory at finite $r_c$:
\be\label{newd}
Z_{EFT}[r_c; \gamma_{ij}, J] = 
Z_{grav}[ g^0_{ij} = r_c^2 \gamma_{ij},
\phi_0 = r_c^{\Delta  - d} J ] \ .
\ee
It is not clear that the right-hand side always makes sense, even classically, since Dirichlet boundary conditions for gravity are problematic (e.g. \cite{Avramidi:1997sh}). Nor is it guaranteed that the QFT on the left really exists. We will simply take the assumption \eqref{newd} as our starting point, and explore whether it leads to a reasonable prescription. We will see in several examples that it does make sense, at least perturbatively about a background, and that the QFT can be constructed as a deformation of the original CFT. 

The deformation involves operators inserted at coincident points.  In general, this would be problematic and require a careful definition of the composite operator. However, at large $N$, we can simply define this operator by normal ordering, in the sense of discarding self-contractions. This is the procedure that will reproduce semiclassical physics in the bulk and is what we adopt here. Equivalently,  ${\cal O}(x)^2$ is defined to be the leading non-identity operator in the ${\cal O}(x) {\cal O}(y)$ OPE that is not suppressed in the $1/N$ expansion.

The rescaling of the sources in \eqref{newd} is natural in the CFT limit, but may not be the most natural choice far from the boundary. Since we will keep track of the full nonlinear source dependence, this is just a change of variables that does not affect the physics. The choice of counterterms at finite cutoff is ambiguous; in particular, to connect to the usual CFT answers as the coupling goes to zero, any counterterms -- not necessarily even local or Lorentz-invariant -- can be added as long as they vanish in the zero coupling limit. As we will see in sections \ref{s:scalar}-\ref{deriv}, the choice of counterterms affects the flow that is derived in the dual field theory. Different choices will lead to different flows in the dual EFT, which by design will have been constructed to match bulk physics at finite cutoff. We will always make the simplest choice of only including the usual holographic counterterms.

The boundary theory on the left of \eqref{newd} is labeled an effective field theory because it has irrelevant operators, and therefore  will not make sense at high enough energy. It is defined in conformal perturbation theory as a CFT plus irrelevant operators. We do not provide a nonperturbative definition of the theory, although we will see that certain quantities -- like the energy spectrum at finite volume and correlation functions -- can be formally calculated at finite $r_c$. 

In the next two sections, the goal is to systematically derive the EFT as a deformation of the original CFT. 

Throughout the paper, we work classically in the bulk, and to leading order in $1/N$ in the boundary. Our notation is as follows:
\begin{align*}
\mbox{Bulk coordinates:} \qquad &(r, x)\\
\mbox{Bulk spacetime metric:} \qquad &g_{\mu\nu}\\
\mbox{Induced metric at }r=r_c: \qquad &g^0_{ij}(x) = g_{ij}(r_c, x)\\
\mbox{CFT metric:} \qquad &\gamma_{ij} = r_c^{-2} g^0_{ij}\\
\mbox{Bulk scalar field:}\qquad &\phi\\
\mbox{Boundary value:}  \qquad &\phi_0(x) = \phi(r_c,x)\\
\mbox{CFT source:} \qquad &J = r_c^{d-\Delta}\phi_0\\
\mbox{Bulk on-shell action:}  \qquad &W[g^0, \phi_0]\\
\mbox{Bulk Brown-York tensor:} \qquad &\tbar_{ij}\\
\mbox{Boundary stress tensor:} \qquad &T_{ij} = r_c^{d-2}\tbar_{ij}
\end{align*}
Various sign conventions are in the appendix.

\section{Scalar example}\label{s:scalar}

We first consider the case where gravity is decoupled, and the bulk theory consists of just a scalar field $\phi$. This section serves to illustrate the methods, including differences from the standard holographic RG, but otherwise stands alone from the rest of the paper and can be skipped. The final answer is equivalent to results found in e.g. \cite{Heemskerk:2010hk, Faulkner:2010jy, Kraus:2018xrn, Mandal:2016rmt}, but our approach is to add nonlinear source dependence to the classical action of the EFT. This gives a local prescription in the boundary theory, in contrast to the scalar discussion in \cite{Kraus:2018xrn}, which was phrased in terms of the non-local effective action.

\subsection{Flow equation of the dual EFT}

Classically, the bulk path integral is computed by the on-shell action, $W[r_c; \phi_0(x)]$. The flow of this functional is governed by the Hamilton-Jacobi equation,
\be\label{hjscalar}
\frac{\p }{\p r_c} W[r_c; \phi_0] = -H[\phi_0, \frac{\delta W}{\delta \phi_0} ]\,,
\ee
where $H[\phi, \pi]$ is the scalar Hamiltonian for evolution in the $r$ direction.  To derive the EFT at finite cutoff, we write $Z_{grav} = e^{-W}$,  apply the flow equation to the dictionary \eqref{newd}, then translate back to the field theory:
\begin{align}
\frac{d}{d r_c} Z_{EFT}[r_c; J=r_c^{d-\Delta}\phi_0] 
&= H[\phi_0, -\frac{\delta}{\delta\phi_0}]e^{-W[r_c; \phi_0]}\\
&= H[\phi_0, -\frac{\delta}{\delta\phi_0}] Z_{EFT}[r_c; J=r_c^{d-\Delta}\phi_0] \notag
\end{align}
This is now written as a total derivative, because $\phi_0$ is fixed but $J$ is not. (Second variations $\frac{\delta^2W}{\delta \phi_0^2}$ drop out in the classical limit, reproducing \eqref{hjscalar}.)
Next, bring the Hamiltonian inside the EFT path integral to obtain
\begin{align}\label{flowinpi}
\frac{d}{d r_c} Z_{EFT}&[r_c; J=r_c^{d-\Delta}\phi_0]
\\&= \int D\varphi H[\phi_0,  -r_c^{d-\Delta}\sqrt{\gamma}{\cal O}] \exp\left( -S_{EFT}(r_c, J; \varphi) + \int d^d x\sqrt{\gamma}{\cal O}\phi_0 r_c^{d-\Delta} \right) \ .\notag
\end{align}
$\varphi$ denotes the fields in the boundary theory. Equating this with  $\frac{d}{dr_c}Z_{EFT} = \int D\varphi \frac{d}{dr_c} e^{-S_{EFT}+\int \sqrt{\gamma}OJ}$ gives the flow equation for the EFT:
\be\label{scalarflow}
\frac{d}{dr_c} S_{EFT}  = -H[r_c^{\Delta-d}J, -r_c^{d-\Delta}\sqrt{\gamma}{\cal O}] + \frac{d-\Delta}{r_c}\int d^d x \sqrt{\gamma}J {\cal O} \ .
\ee
It is convenient to absorb the source term into the action (and not write the $\varphi$ dependence explicitly),
\be\label{shat}
\hat{S}_{EFT}(r_c,J) = S_{EFT}(r_c, J) - \int d^d x \sqrt{\gamma}J {\cal O} \ .
\ee
Then the flow equation takes the form $\frac{d}{dr_c} \hat{S}(r_c, J(r_c)) = - H$. This derivative is taken at fixed value of the bulk boundary condition $\phi_0$, so with $J'(r_c)=\frac{d-\Delta}{r_c}J(r_c)$. For EFT calculations, it is more natural to define the flow in terms of the partial derivative at fixed $J$,
\begin{align}\label{scalarflowp}
\frac{\p}{\p r_c} \hat{S}_{EFT}(r_c,J) &= - H[r_c^{\Delta-d}J, -r_c^{d-\Delta}\sqrt{\gamma}{\cal O}] + \frac{d-\Delta}{r_c}\int d^dx \sqrt{\gamma}J{\cal O}\,.
\end{align}
This is the final result for the scalar. At each step along the flow, the operator ${\cal O}$ must be redefined according to 
\be
{\cal O} = -\f{1}{\sqrt{\gamma}}\frac{\delta }{\delta J}\hat{S}_{EFT} \ .
\ee
Therefore \eqref{scalarflowp}, with the latter relation plugged in, should be viewed as a functional PDE for $\hat{S}$, similar to the Hamilton-Jacobi equation. The difference is that \eqref{scalarflowp} defines the flow of a local action on the boundary, whereas the Hamilton-Jacobi equation \eqref{hjscalar} defines the flow of the non-local, on-shell action in the bulk.

\subsection{Free massive scalar}

To make this formalism explicit, consider a free, massive scalar field in the bulk, 
\be\label{scalaraction}
S_{bulk} = \frac{1}{2} \int d^{d+1}x \sqrt{g}\left( (\del\phi)^2 +  m^2 \phi^2\right)  + \frac{d-\Delta}{2} \int_{\p M} d^d x \sqrt{g^0}\phi_0^2\ ,
\ee
with $m^2 = \Delta(\Delta-d)$, in vacuum AdS, $ds^2 = \frac{dr^2}{r^2} + r^2 dx^2$. The boundary counterterm is added to cancel the leading divergence in the action. 
The radial Hamiltonian is 
\begin{align}\label{scalarhj}
H[\phi_0,\pi] &= \frac{1}{2}\int d^d x N(r_c)\left( 
\frac{ \pi^2}{\sqrt{g^0}} 
- 2(d-\Delta) \pi \phi_0
- \sqrt{g^0} (\p\phi_0)^2\right) \ .
\end{align}
The counterterm has been included by integrating by parts to write it as a bulk term in the action.
The EFT dual to this theory at finite cutoff is defined by the flow equation 
\eqref{scalarflowp}, which together with our dictionary gives
\be\label{scalardef}
\frac{\p}{\p r_c} \hat{S}_{EFT}(r_c,J) = \frac{1}{2}\int d^dx \left(- r_c^{d-2\Delta-1}{\cal O}^2 +  r_c^{2\Delta-d-3}(\p J)^2 \right).
\ee
This defines the corresponding deformation of the CFT, where $r_c$ is now viewed as a dimensionful coupling constant of the EFT.\footnote{Note that the linear term in \eqref{scalarflow} canceled, due to the counterterm. In principle we should include additional counterterms to cancel all of the divergences, which would produce additional terms in the deformation; however this is unnecessary to compute the two-point function. We will include the full set of counterterms in the gravitational case.}

\subsection{Scalar correlators}\label{scalarcorr}

Now we will demonstrate how to obtain bulk correlation functions at finite cutoff, using the boundary theory defined by \eqref{scalardef}. The background terms play a crucial role. Of course this check is guaranteed to succeed, because by design, the bulk and boundary correlators obey the same flow equation.

In the bulk, the on-shell action is quadratic, so for any value of $r_c$ it takes the form
\be\label{actionn}
W[r_c, \phi_0(k)] = \frac{1}{2} \int \frac{d^dk}{(2\pi)^d}\phi_0(k)\phi_0(-k)F(r_c; k) \ .
\ee
$F$ is calculated by solving the wave equation with Dirichlet boundary conditions and plugging into the action. This is a standard exercise that leads to
\be\label{sbessels}
F(r_c;k) = r_c^d (d-\Delta)  + r_c^{d+1} \left. \frac{\p_r( r^{-d/2}K_\nu(k/r))}{ r_c^{-d/2}K_\nu(k/r)}\right|_{r=r_c} \,
\ee
where $\nu = \sqrt{d^2/4+m^2}$. According to our dictionary \eqref{newd}, this gives the boundary two-point function $G(r_c; k)=-r_c^{2(\Delta-d)}F(r_c; k)$. The function $G$ is defined through the correlator in momentum space:
\be 
\braket{\mathcal{O}(\mathbf{k})\mathcal{O}(\mathbf{k}')} = (2\pi)^d \d(\mathbf{k}+\mathbf{k}')G(r_c;k).
\ee
Now we will reproduce this from a boundary calculation. The flow equation \eqref{scalarflowp} implies for the two-point function
\begin{align}
 \frac{\p}{\p r_c} \left(\f{1}{\sqrt{\gamma}}\frac{\delta}{\delta J}\right)\left(\f{1}{\sqrt{\gamma}}\frac{\delta}{\delta J}\right) &\log Z_{EFT}\Big|_{J=0}
= -\frac{\delta}{\delta J}\frac{\delta}{\delta J} \left\langle  \frac{\p}{\p r_c} \hat{S}_{EFT} \right\rangle\Bigg|_{J=0}
\\&=-\int d^dx \frac{\delta}{\delta J}\frac{\delta}{\delta J} \left\langle \half r_c^{2\Delta-d-3}(\p J)^2 - \half r_c^{d-2\Delta-1}O^2 \right\rangle\Big|_{J=0}\notag
\end{align}
Therefore
\be\label{scalartf}
\frac{d}{dr_c} G =  -k^2 r_c^{2\Delta-d-3}+ r_c^{d-2\Delta-1}G^2 \ .
\ee
In the last term, we have invoked the large-$N$ normal-ordering procedure discussed below \eqref{newd} to write
\be
\frac{\delta}{\delta J(x_1)} \frac{\delta}{\delta J(x_2)} \langle {\cal O}(x)^2\rangle\Big|_{J=0} = 2\frac{\delta \langle {\cal O}(x)\rangle}{\delta J(x_1)}\frac{\delta \langle {\cal O}(x)\rangle}{\delta J(x_2)} \Bigg|_{J=0}=2G(x-x_1)G(x-x_2) \,,
\ee
where $G(x-x_1)= \langle \mathcal{O}(x)\mathcal{O}(x_1)\rangle$. The solution of \eqref{scalartf}, if we impose the CFT form at $r_c\to\infty$, is $G(r_c;k)=-r_c^{2(\Delta-d)}F(r_c; k)$, with $F(r_c;k)$ given by \eqref{sbessels}.

Notice that the correlator does not flow for $\Delta = (d+1)/2$, which has a natural explanation in the bulk. This value of the scaling dimension corresponds to a conformally coupled scalar. Weyl invariance then allows us to rescale $r_c$ (or more accurately $\ell_{AdS}$, but this has the same effect since we have set $\ell_{AdS} = 1$). Notice that Weyl invariance is crucial; the argument does not work for massless fields in the bulk (unless $d=1$), and it is easily checked that the correlator for such fields has a nontrivial flow. We will see this feature again in section \ref{u1} when we compute the flow of the two-point function of a Maxwell gauge field, where we will find that the correlator does not flow when $d=3$.

\section{Deriving the deformation with gravity}\label{deriv}

Now we turn to the general case of gravity coupled to matter. The deformation can be derived in two different ways. The first is to find an equation for the trace of the renormalized Brown-York stress tensor, just as was done in two dimensions by \cite{Kraus:2018xrn}. The second derivation is more directly analogous to the scalar example in section \ref{s:scalar}, and follows from the observation that the partition function of the EFT on a radial slice has to be a solution to the radial Wheeler-DeWitt equation in order to describe gravitational physics. 

The two derivations are essentially equivalent, but offer different perspectives. We will describe both.

\subsection{Trace equation}

Consider a Euclidean radial slicing of the form
\be\label{slicing}
ds^2 = \frac{dr^2}{r^2} + g_{ij}(r,x)dx^i dx^j,
\ee
where $g^0_{ij}(x) \equiv g_{ij}(r_c,x)$ is the metric on the cutoff surface. The renormalized Brown-York stress tensor is \cite{Balasubramanian1999} 
\be
\tbar_{ij} = \frac{1}{8\pi G}\left( K_{ij} - K g_{ij}^0 + (d-1)g_{ij}^0 \right)   - a_d \widetilde{C}_{ij} \ ,
\ee 
where $a_d$ is a constant. See the appendix for conventions. We have separated the counterterms into two pieces:  The counterterm $\int_{\p M} \sqrt{g^0}$ gives the $g^0_{ij}$ contribution above, and the curvature-dependent counterterms define the quantity $\widetilde{C}_{ij}[g_{ij}^0]$. Tildes are reserved for bulk quantities which  will appear in our final deformation; they need to be appropriately rescaled to translate into EFT variables.
This stress tensor satisfies
\be\label{traceEq}
\tbar_i^i + a_d \widetilde{C}_i^i = -4\pi G  \left[ (\tbar_{ij} + a_d\widetilde{C}_{ij})^2 - \frac{1}{d-1}(\tbar_i^i + a_d \widetilde{C}_i^i)^2 \right] - \frac{\widetilde{R}}{16\pi G} +\widetilde{t}^r_r \,,
\ee
where $\tilde{t}_{ij}$ is the matter stress tensor. This equation can be derived by plugging in the expression for $\tbar_{ij}$ in the right-hand side and using the Hamiltonian constraint (for the radial slicing) in the bulk:
\be
K^2 - K_{ij}K^{ij} - d(d-1) -\widetilde{R} + 16 \pi G \widetilde{t}^r_r = 0\,.
\ee
From this equation for the stress tensor, we can infer the deformation in the field theory, in the sense of a flow equation. We will temporarily drop the matter term $\widetilde{t}_{ij}$ to derive the flow for pure Einstein gravity. We write the deformation of the classical action in terms of a local operator $X$ as
\be\label{defDeform}
\frac{\p S_{EFT}}{\p\l} = \int d^d x \sqrt{\gamma}X\,,
\ee
with $\l$ a dimensionful parameter that governs the size of the deformation.

In a theory with only one dimensionful scale $\lambda$, invariance under a change of units implies for the effective action 
\be\label{trace}
\l \frac{\p W}{\p \l} = \frac{1}{\D_\lambda}\int d^d x \sqrt{\g} \langle T_i^i\rangle\,,
\ee
with $\D_\lambda$ the mass dimension of $\l$. Combining \eqref{trace} and \eqref{defDeform} with the \emph{bulk} trace relation \eqref{traceEq} suggests the deformation 
\be\label{easyx}
X = (T_{ij}+a_dr_c^{d-2} \widetilde{C}_{ij})^2 - \frac{1}{d-1}(T_i^i+a_d r_c^{d-2} \widetilde{C}_i^i)^2 - \frac{r_c^d}{d\l}\left(\widetilde{t}^r_r - \frac{\widetilde{R}}{16\pi G} - a_d \widetilde{C}_i^i\right)   \ .
\ee
This is a field theory equation, so $T_{ij}$ is the field theory stress tensor, and indices are raised with $\gamma^{ij}$. It was obtained from \eqref{traceEq} by replacing bulk with boundary quantities, $g_{ij}^0 = r_c^2\gamma_{ij}$, and $\tbar_{ij} = r_c^{2-d}T_{ij}$. Other tilded quantities must also be rescaled, which we will do when considering explicit examples. 

With this choice of deformation operator, we have $\Delta_\lambda = -d$ and the relation between the boundary coupling $\lambda$ and bulk radial cutoff $r_c$ is
\be \label{genlam}
\lambda = \frac{4\pi G }{d\, r_c^d} \ .
\ee
For a four or five dimensional bulk
\be 
\widetilde{C}_{ij} = \widetilde{G}_{ij}=G_{ij} , \qquad a_d = \frac{1}{8\pi G(d-2)} \ ,
\ee
with $\widetilde{G}_{ij}$ the Einstein tensor for $g_{ij}^0$ and $G_{ij}$ the Einstein tensor for $\g_{ij}$. In general dimensions for a flat metric $\gamma_{ij}$ we have $\widetilde{C}_{ij} = 0$.

There are some subtleties in this argument. The first is the issue of anomalies. The expression \eqref{easyx} includes terms built entirely from background fields. The $O(r_c^0)$ background-only terms in $r_c \p_{r_c} W$, which occur only in even $d$, give precisely the conformal anomaly (since they correspond to $\log r_c$ terms in $W$). The interpretation is that we are implicitly measuring the UV cutoff in units of $\lambda$, so the UV cutoff changes along the flow, and this contributes to \eqref{defDeform} via the Weyl anomaly. 
In other words, $S_{EFT}$ must be regulated, and the effect of the regulator has been included in \eqref{easyx}. This will be clear in the even-dimensional examples below.

Also, in \eqref{easyx}, we have assumed that there are no additional contributions, beyond the trace anomaly, from renormalization. This is not obvious, and will only be justified \textit{a posteriori} by comparison to the bulk. Finally, the composite operators in \eqref{easyx} must be regulated somehow. In the 2d case, it turned out that the regulator was unnecessary, due to the factorization property \cite{Zamolodchikov:2004ce}; in higher dimensions, we use the large-$N$ normal ordering procedure discussed in section \ref{ss:dictionary}. 

Let us now give the explicit form of the deformation $X$ in two, three, and four boundary dimensions without bulk matter. In two and four dimensions we will make the contribution of the trace anomaly manifest.

\subsubsection*{Deformation in $d=2$}

In two boundary dimensions, the deformation was already derived in the references cited above, but we will give it for completeness. The trace anomaly is
\be\label{anomaly2d}
\mathcal{A} \equiv \langle T^i_i\rangle_{CFT}  = -\frac{c}{24\pi}R = -\frac{r_c^2}{16\pi G} \widetilde{R}\, ,
\ee
where we used $c = 3/2G$ and $R = r_c^2 \widetilde{R}$. In $d=2$ we have $\widetilde{C}_{ij}=0$ since all curvature counterterms are absent. Combining this with \eqref{anomaly2d} we deduce that the deformation is
\be 
X = T_{ij}T^{ij} - (T_i^i)^2 - \frac{1}{2\lambda}{\cal A}\, .
\ee
The first two terms are often denoted in terms of $T\bar{T} \equiv 1/8\left(T_{ij}T^{ij} - (T_i^i)^2\right)$. As discussed above, the total deformation $X$ includes both the explicit deformation of the EFT Lagrangian by the operator $\delta {\cal L} = T\bar{T} \delta \lambda$, and the contribution from the Weyl anomaly as we rescale the UV cutoff.

\subsubsection*{Deformation in $d=3$}
For a three dimensional boundary, there is no trace anomaly, so the deformation is 
\be
X = (T_{ij} + a_3 r_c G_{ij})^2 - \frac{1}{2}(T_i^i + a_3 r_c G_i^i)^2 \ .
\ee
In this equation, $a_3 r_c$ has to be expressed in terms of boundary data:
\be 
a_3 r_c = \frac{1}{6^{1/3}}\left(\frac{1}{8\pi G}\right)^{2/3}\l^{-1/3} = \a_3\l^{-1/3} \ .
\ee
Here $\a_3$ is a function of $N$ on the boundary, but is independent of $\l$.
The deformation of the boundary theory is thus
\be\label{3dDeform}
X = \left(T_{ij} + \frac{\a_3}{\l^{1/3}}G_{ij}\right)^2 - \frac{1}{2}\left(T_i^i + \frac{\a_3}{\l^{1/3}} G_i^i\right)^2 
\ee
Note that despite the inverse powers of $\lambda$, the CFT limit $\lambda \to 0$ is regular, since the first order deformation is $\delta {\cal L} = \lambda X$.

\subsubsection*{Deformation in $d=4$}

In a four dimensional boundary theory, the trace anomaly for a theory dual to Einstein gravity is 
\be\label{anomaly}
\mathcal{A} = -\frac{C_T}{8\pi}\left(G_{ij}G^{ij} - \frac{1}{3}(G_i^i)^2\right) \ ,
\ee
with $C_T = \frac{1}{8G}$ \cite{Skenderis1998}. Therefore the deformation \eqref{easyx} may be written
\be\label{4dDeform}
X = T_{ij}T^{ij} - \frac{1}{3}(T_i^i)^2 + 2a_4 r_c^2\left(G^{ij}T_{ij} - \frac{1}{3}G_i^i T_j^j \right) - \frac{1}{4\lambda}{\cal A} 
\ee
with $a_4 r_c^2= \a_4 \l^{-1/2}$. As before, $\a_4$ is fixed in terms of $1/G$.

\subsubsection*{Comments on the flow equation}

Note that in going infinitesimally from $r_c \to r_c  + \delta r_c$, the \textit{deformed} $T_{ij}$, at the value $r_c$, must be used on the right-hand side of the flow equation.  This means that, like the Hamilton-Jacobi equation, it must be viewed as a functional equation for ${S}$, with $T_{ij} = -\frac{2}{\sqrt{\gamma}}\frac{ \delta S}{\delta \gamma^{ij}}$. The difference, however, is that this defines the flow of a local functional -- the EFT action -- whereas the Hamilton-Jacobi equation governs the flow of the bulk on-shell action.

There is one last subtlety to address in the meaning of the flow equation \eqref{trace}. This is written as a partial derivative $\frac{\partial}{\partial \lambda}$ because the EFT metric, and other sources if present, are  held fixed. Ultimately, however, the dictionary \eqref{newd} equates the bulk theory to a boundary theory where the boundary metric $\gamma_{ij}=r_c^{-2}g_{ij}^0$ may itself be a function of $r_c$. The only time this will occur in our examples is when we consider bulk black hole geometries to compute the deformed energy spectrum. In this case, $\gamma_{00}$ is $r_c$-dependent but the metric is diffeomorphic to the original, undeformed metric. Thus we can compare quantities computed in the deformed and undeformed metrics simply by a coordinate rescaling at the end of the calculation. But if the intrinsic geometry of the boundary changes along the flow, then the bulk is dual to $S_{EFT}[\lambda, \gamma_{ij}(\lambda)]$, which includes an additional term $\half T^{ij} \p_{r_c}\gamma_{ij}$ in the total flow equation for $\frac{d}{d\lambda}S_{EFT}$.

Sources for irrelevant operators can induce a large backreaction, destroying the AdS boundary. This is not a problem here because of the finite radial cutoff --- the backreaction is assumed to be small where the flow starts, at some large but finite $r_c$, and any problems with the AdS asymptotics occur only for $r > r_c$. This is similar to the situation in the boundary EFT, where the CFT deformed by an irrelevant operator is sensible below the cutoff set by the mass scale of the deformation.

\subsection{Wheeler-DeWitt method}
Another perspective on this derivation is provided by the Wheeler-DeWitt equation. This is closer to the scalar derivation in section \ref{s:scalar}, where we translated the bulk Hamilton-Jacobi equation into a deformation of the dual EFT. We will also include matter sources in this discussion.

Without gravity, the Hamilton-Jacobi equation governs the flow of the on-shell action $\f{\p}{\p r_c} W[r_c; \phi_0]$, with boundary sources held fixed.  But once gravity is included, this equation is trivial, because the on-shell action is no longer an explicit function of the cutoff $r_c$ --- explicit dependence on $r_c$ is a pure diffeomorphism, so does not affect the value of the action. Instead, the action depends on the cutoff only via the induced metric, $W = W[g^0_{ij}, \phi_0]$. To keep track of this, we define the flow with boundary values of non-metric fields $\phi_0(x)$ held fixed, but make the induced metric a function of the cutoff, $g^0_{ij} = g^0_{ij}(r_c, x)$.

Assume the bulk metric takes the form \eqref{icoord}.
The renormalized on-shell action (i.e. including holographic counterterms) obeys
\begin{align}\label{tsflow}
\frac{d}{d r_c} W[g^0_{ij}(r_c,x), \phi_0]
= \frac{1}{2} \int d^d x \sqrt{g^0}\ \tbar^{ij} \p_{r_c} g^0_{ij}
\end{align}
where $\tbar^{ij}$ is the renormalized Brown-York stress tensor, since it is obtained by varying the renormalized on-shell action. Writing $g^0_{ij} = r_c^2 \gamma_{ij}$, this becomes
\be\label{tsflow2}
r_c \frac{d}{dr_c} W[g^0_{ij}(r_c,x), \phi_0] = \int d^d x \sqrt{g^0} \tbar^i_{\ i} + \frac{r_c^3}{2} \int d^d x\sqrt{g^0} \tbar^{ij}\p_{r_c} \gamma_{ij} \ .
\ee
At this point, this separation of the trace is somewhat arbitrary, but useful, as we will see. This is turned into a flow equation by substituting the Hamiltonian constraint into the first term. In general, this constraint can be written 
\be\label{cons}
\tbar^{i}_{\ i} = \Theta \ ,
\ee
where $\Theta$ is built from both $\tbar_{ij}$ and the matter fields. Although \eqref{tsflow2} has no dynamics as written, once we replace $\tbar^{i}_{\ i} \to  \Theta$, it becomes the classical Wheeler-DeWitt equation, which encodes the dynamical equations of the classical bulk theory, and is the gravitational analogue of the Hamilton-Jacobi equation.

Now we repeat the argument used in the scalar case to derive the flow equation in the dual EFT. First, write $\Theta$ in terms of the canonical data:
\be
\Theta = \Theta[g^0_{ij}, p_{ij}, \phi_0, \pi] \ ,
\ee
where $p_{ij} \equiv \sqrt{g^0}\tbar_{ij}$ and $\pi$ is the momentum conjugate to $\phi$. Translating \eqref{tsflow} into the language of the boundary field theory, we have
\begin{align}
r_c \frac{d}{dr_c} &Z_{EFT}[r_c; \gamma_{ij}=r_c^{-2}g^0_{ij}, J=r_c^{d-\Delta}\phi_0] 
\\&=  -\left( \int d^d x \sqrt{g^0}\,\Theta[g^0_{ij}, -\frac{2\delta}{\delta g^{0ij}}, \phi_0, -\frac{ \delta}{\delta \phi_0}] + r_c\int d^d x\, \p_{r_c}\gamma_{ij} \frac{\delta}{\delta \gamma_{ij}}\right) Z_{EFT}\,.
\end{align}
Pulling this inside the EFT path integral, as we did for the scalar around \eqref{flowinpi}, gives (up to the anomaly)
\begin{align}\label{flowg}
r_c \frac{d}{dr_c} \hat{S}_{EFT} &= r_c^d \int d^dx \sqrt{\gamma} \,\Theta[r_c^2 \gamma_{ij}, r_c^2 \sqrt{\gamma} T_{ij}, r_c^{\Delta-d}J, -r_c^{d - \Delta}{\cal O}] 
\\& \quad + \frac{r_c}{2}\int d^d x \sqrt{\gamma}\,T^{ij}\p_{r_c}\gamma_{ij}\,. \notag
\end{align}
We have included the full source dependence in $\hat{S}_{EFT}$, defined e.g. as in \eqref{shat} for a scalar source with corresponding generalizations for other fields. As discussed above, counterterms are also included, so the flow includes the contribution of the trace anomaly.

\subsection{The final prescription}
To recap, the general answer is as follows. The deformation of the boundary effective field theory is given by the flow equation 
\be\label{partialflow}
r_c \frac{\partial}{\partial r_c} S_{EFT} = r_c^d
 \int d^dx \sqrt{\gamma}\, \Theta  \,.
\ee
 $\Theta$ is the right-hand side of the constraint equation \eqref{cons}, with the rescalings appropriate to translate from bulk to boundary variables,
\be\label{replace}
g_{ij}^0 \to r_c^2 \gamma_{ij} , \quad
\tbar_{ij} \to r_c^{2-d}T_{ij} , \quad
\phi_0 \to r_c^{\Delta-d}J , \quad
\pi \to \sqrt{\gamma}r_c^{d-\Delta} {\cal O} \ .
\ee
The partial derivative in \eqref{partialflow} is taken with $\gamma_{ij}, J$ held fixed -- but to match the bulk, the sources and background metric must also be modified along the flow according to \eqref{flowg}. The rescalings of $J, {\cal O}$ for spin-$L$ fields have additional factors of $r_c^L$. For bulk $p$-forms, which we will consider for $p=1$ in the next subsection, we have
\be\label{pforms}
\phi^0_{\mu_1\cdots \mu_p} \rightarrow r_c^{\Delta-d+p}J_{\mu_1 \cdots \mu_p}\,,\qquad \pi^{\mu_1 \cdots \mu_p}\rightarrow \sqrt{\gamma}r_c^{d-\Delta-p}\mathcal{O}^{\mu_1\cdots\mu_p}\,.
\ee

\subsection{Matter contributions and the $U(1)$ case}\label{addingMatter}

Matter is automatically included in the prescription \eqref{partialflow}, simply by including the matter Hamiltonian $\tilde{t}^r_r$ on the right-hand side of the  constraint equation \eqref{cons}. This reproduces, for example, the scalar results in section \ref{s:scalar}, upon sending $G \to 0$ with the matter action held fixed.

Another interesting case is a $U(1)$ gauge field $A_\mu$ in the bulk, dual to a conserved $U(1)$ current in the boundary field theory. The Dirichlet boundary condition in AdS/CFT fixes the non-normalizable mode of $A_{\mu}$, which means fixing the chemical potential $\mu$ of the boundary field theory.

For a Maxwell field in the bulk the Euclidean Lagrangian is given by
\be 
\mathcal{L}_m = \frac{1}{4e^2}F_{\mu\nu}F^{\mu\nu}.
\ee
Its stress tensor follows from the usual prescription and the $rr$ component reads
\be 
\tilde{t}_{r}^{r} = \frac{1}{4e^2}F_{ij}F^{ij} - \frac{1}{2e^2}F^{ri}F_{ri}
\ee
To apply our dictionary, we can write this in terms of the canonical momentum $\pi_i$ of the gauge field in the bulk,
\be 
\tilde{t}_r^r = \frac{1}{4e^2}F_{ij}F^{ij} - \frac{e^2}{2}\frac{\pi_i\pi^i}{\left(\sqrt{g^0}\right)^2}.
\ee
The canonical momentum $\pi_i$ and bulk non-normalizable mode $A_{(0)i}$ are related to the boundary operator and source as 
\be 
\pi_i \to \sqrt{\g}r_c^2J_i, \quad A_{(0)i} \to A_{i}.
\ee
In combination with \eqref{replace}, we find
\be 
\tilde{t}_r^r = -\frac{e^2}{2}r_c^{2-2d}J_iJ^i +\frac{1}{4e^2}r_c^{-4}F_{ij}F^{ij},
\ee
where all contractions are done with $\g_{ij}$. The flow of the effective action is thus 
\be \label{deformMatter}
r_c \frac{\partial}{\partial r_c} S_{EFT} = r_c^d \int d^d x \sqrt{\g}\left( \Theta_{\rm grav}  - \frac{e^2}{2}r_c^{2-2d}J_iJ^i +\frac{1}{4e^2}r_c^{-4}F_{ij}F^{ij}  \right).
\ee

For a complete identification of bulk data with boundary data, we have to convert constants such as $e^2$ to boundary data. This quantity has dimension $3-d$ and is related to the coefficient $C_J$ of the two-point function of conserved currents. Specifically, in the field theory on $\mathbb{R}^d$ this two-point function is 
\be\label{JJcorrelator} 
\braket{J_i(x)J_j(y)} = \frac{C_J}{(2\pi)^d}\left(\partial^2 \delta_{ij} - \partial_{i}\partial_{j}\right)\frac{1}{|x-y|^{2(d-2)}},
\ee 
and the relation between $C_J$ and $e^2$ is given by
\be \label{e2CJ}
e^2 = \frac{1}{C_J \a_J},\quad \a_J = \frac{(d-1)\G(d/2)}{2^{d-2}\pi^{d/2}\G(d)}\,.
\ee
This allows us to translate any coefficient in \eqref{deformMatter} to functions of $\l$ and dimensionless numbers that are, as we will see, powers of $N$. For example, rewriting the $r_c$ dependence in terms of $\l$, we find:
\be 
\f{\p S_{EFT}}{\p \l} 	\supset \int d^d x \sqrt{\gamma}\left(\f{\sigma_1}{\l^{2/d}} J_iJ^i - \f{\sigma_2}{\l^{2(d-2)/d}} F_{ij}F^{ij}\right)
\ee
with
\be \label{gammas}
\sigma_1 = \frac{1}{8\pi\a_J C_J}\left(\frac{4\pi}{d}\right)^{2/d} G^{(2-d)/d},\quad \sigma_2 = \f{\a_J C_J}{16\pi}\left(\frac{4 \pi }{d}\right)^{2(d-2)/d}G^{(d-4)/d}\,.
\ee
The Newton constant $G$ is proportional to some power of $N$, so $\sigma_1$ and $\sigma_2$ are fully expressed in terms of boundary data. Said another way, the coefficients can be expressed in terms of the central charges $C_J$ and $C_T$ of the two-point functions of a conserved $U(1)$ current $J_i$ and a conserved spin-two current $T_{ij}$.

\section{Random metrics via Hubbard-Stratonovich}

The deformation can also be understood as coupling to a random background metric. This was explored in $d=2$ in \cite{Cardy:2015xaa,McGough:2016lol}. Here we will show that in general $d$, the radial flow equation for the induced metric --- that is, the bulk Einstein equation for $\gamma_{ij}$ --- is precisely the flow induced by coupling to a random background metric. In this section we assume $\gamma_{ij}$ is flat.

Let us introduce a symmetric two-tensor $h_{ij}$ as our Hubbard-Stratonovich field and rewrite the deformation as
\begin{multline}\label{HS}
\exp\left(-\delta\lambda \int d^d x \sqrt{\gamma} \left(T_{ij}T^{ij} - \frac{1}{d-1}(T^i_i)^2\right) \right) \sim\\
 \int \mathcal{D}h \exp\left( - \frac{1}{16\d\l}\int d^d x \sqrt{\gamma}(h^2 - h_{ij}h^{ij}) + \frac{1}{2}\int d^dx \sqrt{\gamma} h_{ij}T^{ij} \right),
\end{multline}
where $h = h^i_i$. From this rewriting we see that the deformation corresponds to coupling to a  metric perturbation $h_{ij}$, and averaging over $h_{ij}$.
The saddle point equations are 
\be
h \d_{ij} -  h_{ij} -4\d\l T_{ij} = 0\,.  
\ee
Taking the trace of this equation tells us that $(d-1)h = 4\d\l T^i_i$, so
\be\label{hijj}
h_{ij} =  -4\d\l \left(T_{ij} - \frac{T^k_k}{d-1}\g_{ij}\right).
\ee
Assuming a large, classical background stress tensor, this can be interpreted as a change $\delta \gamma_{ij}$ in the effective metric seen by the field theory.

Now let's compare to the bulk. The radial evolution equation for the induced metric on a fixed-$r$ slice is Hamilton's equation, 
\be
\partial_{r_c}g_{ij} = 16\pi G N\left(\pi_{ij} - \frac{\pi^k_k}{d-1}g_{ij}\right),
\ee
where the lapse and canonical momentum are 
\be
N = \frac{1}{r_c} , \qquad \pi_{ij} = \frac{1}{8\pi G}(K_{ij} - K g_{ij}) \ .
\ee
Setting $g_{ij}  = r_c^2 \gamma_{ij}$, this becomes
\be
r_c^3 \frac{\p \gamma_{ij}}{\p r_c} = 16\pi G \left(\tbar_{ij} - \frac{\tbar^k_k}{d-1}\g_{ij}\right) \ .
\ee
Upon rescaling $\tbar_{ij} = r_c^{2-d}T_{ij}$ and using \eqref{genlam}, this agrees with the flow of the effective metric \eqref{hijj}.

At first order in the deformation, the effective metric is 
\be
ds^2 = ds_0^2 - 4 \lambda \langle T_{ij}\rangle dx^i dx^j \ .
\ee
Viewed as a bulk equation for the induced metric, this is the usual dictionary for the boundary stress tensor in terms of subleading terms in the bulk metric. 

Let us compute the propagation speed when $T_{ij}$ is diagonal with components $T_{tt} = \e$, $T_{ii} = \frac{\e}{d-1}$. With this choice, we can focus on a two dimensional plane, say the $(t,x)$ plane, to perform this calculation. In Lorentzian signature, the null geodesics in this plane are
\be 
-dt^2 - 4\l \braket{T_{tt}}dt^2 + dx^2 - 4\l \braket{T_{xx}} dx^2 = 0\,.
\ee
In the small $\l$ limit the propagation speed $v$ is thus
\be
v = 1 + 2\l \e \frac{d}{d-1} + \mathcal{O}(\l^2)\,.
\ee
For the theory on $\mathbb{R}^{d-1}$, $\epsilon \geq 0$ and this speed is superluminal for $\l>0$. However for the theory on e.g. $\mathbb{T}^{d-1}$ with thermal periodicity conditions along the spatial cycles, the vacuum necessarily has $\epsilon < 0$ \cite{Belin:2016yll}, in which case we can have $v>1$ for $\l<0$ as well.

\section{Spectrum}
In this section we will consider the deformed energy spectrum of a large-$N$ CFT under the $T^2$ deformation. Thanks to factorization, we will have a single differential equation that governs all energy levels. We will solve this equation and match the answer to a bulk computation of the energy at finite cutoff of black holes in anti-de Sitter space. We will consider the general case of finite sources for curvature and $U(1)$ charge, which will require considering charged AdS-Reissner-Nordstr\"om black holes with curved horizons.

\subsection{Field theory analysis}
We study field theories on a manifold $\mathbb{R} \times \mathcal{M}^{d-1}$ with metric
\be
ds^2 = d\tau^2 + h_{ab}dx^a dx^b\,.
\ee
The flow defined by $\partial S/\partial \lambda = \int d^d x \sqrt{\gamma} X$  implies the same flow for the Hamiltonian and therefore for the energy levels, $\partial E/\partial \lambda = \int d^{d-1} x \sqrt{\gamma} X$. 
Considering states that preserve the symmetries of $h_{ab}$ and in which large-$N$ factorization holds, and after passing to densities by dropping the spatial volume integrals, we have
\be\label{eneq}
\frac{\partial \e}{\partial \l} = \left< T_{ij} + \frac{\a_d}{\l^{\frac{d-2}{d}}} G_{ij}\right >^2 - \frac{1}{d-1}\left<T_{i}^{i} + \frac{\a_d}{\l^{\frac{d-2}{d}}} G_{i}^{i}\right>^2  \ .
\ee 
Equation \eqref{eneq} is valid for any $d$ if $\mathcal{M}^{d-1}$ is flat, and for $d=3,4$ when $\mathcal{M}^{d-1}$ is arbitrary. This is the main object to study in the field theory as it will determine the deformed spectrum of our states of interest as a function of the deformation parameter $\l$. We will now solve this differential equation for various backgrounds. 

Before discussing the deformation in full generality, let us focus on the simplest case in which the CFT is living on a square torus $\mathbb{T}^{d-1}$. For this background the Einstein tensor $G_{ij}$ vanishes and moreover there are no trace anomalies. Let us assume that the states do not carry any momentum so that the stress tensor is diagonal in these states. The diagonal components of the stress tensor are given in terms of the energy density as 
\be
\langle T_{\t\t}\rangle = \e,\quad \langle T_{aa}\rangle = \frac{1}{\prod_{b\neq a} L_j} \frac{d (\e \prod_b L_b)}{dL_a}\,.
\ee
For a square torus the stress tensor is diagonal with equal spatial components, and the differential equation becomes
\be
\frac{\partial \e}{\partial \l} = \frac{d-2}{d-1}\e^2 - \frac{2\e}{(d-1)L^{d-2}}\partial_L (L^{d-1} \e).\\
\ee 
Solutions to this differential equation in terms of the energy $E = \e L^{d-1}$ are given by 
\be
E = \frac{(d-1)L^{d-1}}{2d \l}\left(1 - \sqrt{1-\frac{4d\l}{d-1}\frac{M}{L^d}}\right),\label{cften}
\ee
where the undetermined constant was fixed by requiring that as $\l \to 0$ we obtain the energy in the undeformed theory $E^{0} = M/L$. 

At $E^0_{\text{max}} \defeq (d-1)L^{d-1}/(4d\l)$ the energy levels exhibit a ``square-root singularity" and become complex.  For the theory with $\l>0$, which we will argue is dual to the finite cutoff theory in AdS, this affects an infinite number of positive energy states. This suggests a maximum energy and hence a sharp UV cutoff. In the bulk description, it affects all states with energies for which a black hole of the given energy would not fit inside the cutoff, i.e. its horizon radius is bigger than $r_c$. For the theory with $\l<0$, this can only affect negative energy states in the spectrum, which will necessarily exist if e.g. we pick thermal periodicity conditions along the spatial cycles \cite{Belin:2016yll}. While the theory with $\l>0$ has complex energy states for any $\l$ and $L$, the theory with $\l<0$ can only have complex energy states for sufficiently large $-\l/L^d$. 

We now consider the general case, where we will solve the differential equation for the energy levels with finite $U(1)$ charge density on $S^{d-1}$ ($k=1$) or $\mathcal{H}^{d-1}$ ($k=-1$). The metric is given by
\be
ds^2 = d\tau^2 + R^2 d\Sigma^2_{d-1},
\ee
with $R^2 d\Sigma^2_{d-1}$ the metric on an $S^{d-1}$ or $\mathcal{H}^{d-1}$ with radius $R$ and volume $R^{d-1}V_{d-1}$. (The flat slicing case treated above is captured by taking the flat metric on $\Sigma_{d-1}$, which in the below equations will mean setting $k=0$, $V_{d-1} = 1$ and $R= L$.)
For simplicity, let us restrict to states that preserve the spatial symmetries. This means that the stress tensor is given by 
\be
T_{\tau\tau} = \e, \quad T_{aa} = h_{aa} \f{1}{(d-1)R^{d}}\p_R\left(R^{d-1}\e\right) \ . 
\ee
In the presence of finite $U(1)$ charge density the deformation was shown in section \ref{addingMatter} to be given by 
\be 
X = \left(T_{ij} + \frac{\a_d}{\l^{\frac{d-2}{d}}} G_{ij}\right)^2 - \frac{1}{d-1}\left(T_{i}^{i} + \frac{\a_d}{\l^{\frac{d-2}{d}}} G_{i}^{i}\right)^2+ \frac{\sigma_1}{\l^{2/d}}J^{i}J_{i} - \frac{\sigma_2}{\l^{\f{2(d-2)}{d}}}F_{ij}F^{ij}\,,
\ee
where $\sigma_i$ are dimensionless constants given in \eqref{gammas}. For simplicity, let us study the deforming operator when $A_i$ is independent of field theory coordinates. Following the same logic as above, the flow of the energy levels in the deformed theory are given by 
\be 
\frac{\partial \e}{\partial \l } = \left<T_{ij} + \frac{\a_d}{\l^{\frac{d-2}{d}}} G_{ij}\right>^2 - \frac{1}{d-1}\left<T_{i}^{i} + \frac{\a_d}{\l^{\frac{d-2}{d}}} G_{i}^{i}\right>^2+ \frac{\sigma_1}{\l^{2/d}}\left<J^{i}J_{i}\right> \,.
\ee
Again, by using large-$N$ factorization, we can write all terms as products of one-point functions. We will consider the current one-point functions to vanish when $i \neq 0$, so these states only have a non-zero charge density, which enters into the final term in the flow equation as
\be 
\langle J^iJ_i\rangle = \langle J^i\rangle\langle J_i\rangle =-\left(\frac{Q}{V_{d-1}R^{d-1}}\right)^2\,,
\ee
with $Q$ the dimensionless charge. The differential equation for the energy levels is then given by
\begin{align}
\frac{\partial\e}{\partial\l} = \frac{d-2}{d-1}\left(\e-\frac{(d-1) k \alpha _d}{2 R^2 \lambda ^{1-2/d}}\right)^2&-\frac{2 }{(d-1)R^{d-2}}\frac{\partial \left(R^{d-1} \e\right)}{\partial R} \left(\e-\frac{(d-1) (d-2) k \alpha _d}{2 R^2 \lambda ^{1-2/d}}\right)\nonumber\\\nonumber
\\
&\quad-\frac{(d-3)^2 \left(d^2-3 d+2\right) \alpha _d^2}{4 R^4 \lambda ^{2 \left(1-2/d\right)}}-\frac{\sigma_1 Q^2}{\lambda ^{2/d}R^{2d-2} V_{d-1}^2}\,.
\end{align}
The equation can be simplified by defining an energy variable $x = R^{d-1}\e-(d-1)(d-2)R^{d-3}k\alpha_d/(2\l^{1-2/d})$. It is solved by energy density $\e=E/(R^{d-1}V_{d-1})$, with 
\begin{multline}
E = \f{(d-1)R^{d-1}V_{d-1}}{2d\l}\left(\f{(d-2)d k \l^{2/d}\alpha_d}{R^2}+1-\right.\\
\left.\sqrt{1-\f{4d M \l}{(d-1)V_{d-1}R^d}+\f{2(d-2)d k \l^{2/d}\alpha_d}{R^2}+\frac{4 \sigma_1 d^2 Q^2 \lambda^{2-2/d} }{(d-2) (d-1) R^{2d-2}V_{d-1}^2}}\right).
\end{multline}
For $d=2, 3$ we see that this reduces to the CFT energy $M/R$ as $\l \rightarrow 0$. (For $d=2$ we only consider the chargeless case $Q=0$.) For $d=4$ the CFT limit picks up a Casimir term and becomes $M/R+12 |k|V_3\alpha_4^2/R$. For $d>4$ the limit is singular, reflecting the fact that there are more counterterms we have neglected to include in deriving our deformation. Our bulk calculations will be done with the same set of counterterms, which will result in us matching the $d>4$ cases between bulk and boundary as well. 

\subsection{Bulk analysis}
Having obtained the energy levels in the deformed theory, we now turn to a comparison with the bulk. In the bulk, we want to do a quasi-local energy calculation at a finite radial cutoff for the AdS-Reissner-Nordstr\"om black hole metric with boundary geometry $S^{d-1}$ ($k=1$), $\mathbb{R}^{d-1}$ ($k=0$), or $\mathcal{H}^{d-1}$ ($k=-1$). The topology can be arbitrary and will only enter into the volume $V_{d-1}$. The action for the theory is 
\be
S = -\int d^{d+1}x \sqrt{g}\left(\f{R}{2\kappa^2} - \f{1}{4e^2}F^2+\f{d(d-1)}{2\kappa^2}\right)
\ee
where $\kappa^2 = 8\pi G$ and the gauge coupling is $e$. This has as solution the charged black hole:
\begin{align}
ds^2 &= \left(\frac{k}{R^2}-\frac{r_0}{r^{d-2}}+r^2 +\frac{q^2}{r^{2 d-4}}\right)d\tau^2 + \f{dr^2}{\frac{k}{R^2}-\frac{r_0}{r^{d-2}}+r^2+\frac{q^2}{r^{2 d-4}}}+r^2R^2 d\Sigma_{d-1}^2,\\[0.2cm]
&\hspace{2.5cm}A = \f{ie q}{c\kappa} \left(-\f{1}{r^{d-2}}+\f{1}{r_+^{d-2}}\right) d\tau,\qquad c = \sqrt{\f{d-2}{d-1}}\,,
\end{align}
where $r_+$ is the horizon location and $\Sigma_{d-1}$ has volume $V_{d-1}$ and is a unit sphere, plane, or hyperboloid depending on $k$. 
The conserved mass and dimensionless $U(1)$ charge of the CFT are 
\be\label{MassChargeBB}
M = \f{(d-1)R^{d-1} V_{d-1}}{16 \pi G}\,r_0\,,\qquad Q = \f{\sqrt{(d-1)(d-2)}R^{d-1}V_{d-1}}{e\kappa}\, q\,.
\ee
Using $E = \int \sqrt{\det\, \tilde{h}}\,\widetilde{T}_{\mu\nu} u^\mu u^\nu = \int \sqrt{\det\, \tilde{h}}\,\widetilde{T}_{\tau\tau} g^{\tau\tau}$ for $\tilde{h}_{ab}dx^a dx^b=r^2R^2d\Sigma_{d-1}^2$ the non-radial spatial metric, we find the energy at finite radial cutoff $r_c$ to be
\be 
E = \frac{(d-1)V_{d-1}R^{d-1}r_c^{d-1}}{8\pi G }\left(\frac{k }{2R^2 r_c^2} +1- \sqrt{1-\frac{r_0}{r_c^d}+\frac{k}{R^2 r_c^2}+\frac{q^2}{r_c^{2d-2}}}\right).
\ee
This expression is correct for $d>2$ if $k=0$ and for $d=3, 4$ if $k\neq 0$. The general dimensional answer for $k\neq 0$ can also be obtained but would (further) clutter the equation. 

To translate to field theory we need to apply our dictionary to the quantity $E_{\text{bulk}}=\int \sqrt{\det\, \tilde{h}}\,\widetilde{T}_{\tau\tau} g^{\tau\tau}\rightarrow \int (r_c^{d-1}\sqrt{\det\, h})(r_c^{2-d}T_{\tau\tau})(r_c^{-2}\g^{\tau\tau})=r_c^{-1} E_{\text{bdry}}$. Using the expressions for $\alpha_i$ and $\sigma_i$ in the previous section and identifying
\be\label{mu}
\l = \frac{4\pi G}{dr_c^d}\,,
\ee
we find perfect agreement between $E_{\text{bdry}}$ calculated in this way and $E_{\text{bdry}}$ calculated in the field theory analysis of the previous subsection. 

Note that we calculated the bulk energy by integrating $\tilde{T}_{\mu\nu}u^\mu u^\nu$. Often, the quasilocal energy is defined by integrating $\tilde{T}_{\mu\nu}u^\mu t^\nu$, which differs by a function of $r_c$. Both choices are acceptable, as long as we compare to the correct quantity in the boundary theory. The energy computed from $\tilde{T}_{\mu\nu}u^\mu u^\nu$ is the conserved charge associated to translations by a unit vector, and is therefore equal to the field theory energy in the metric $-dt^2 + \cdots$. The energy defined by integrating $\tilde{T}_{\mu\nu}u^\mu t^\nu$ is equal to the energy of the boundary theory in the metric induced at the cutoff surface, $-\gamma_{tt} dt^2 + \cdots$. Since we are comparing to the EFT energy in Minkowski spacetime with the usual (unit normalized) time coordinate, the correct comparison is to $\tilde{T}_{\mu\nu}u^\mu u^\nu$.


\section{Thermodynamics}
So far, we have only considered the flow of the spectrum of the deformed theory, but there are other quantities that also exhibit a non-trivial flow under the deformation. Two important quantities that reveal some of the intricate features of the $T^2$ deformation are the entropy and speed of sound. We will consider both quantities for the case of the effective theory on a flat background. 

\subsection{Entropy density}

The interpretation of our deformation in terms of a finite cutoff in an AdS bulk requires a particular sign for the deformation, in our conventions $\lambda > 0$. The case $\lambda < 0$ is also interesting to consider. (If matter or sources are present there will be fractional powers of $\lambda$, so the theory needs to be defined more carefully, but here we will only consider the sourceless case without matter.) In this case the deformed energy levels for $E>0$ always remain real, so we can analyze what happens in the deep UV of our system. In the local CFT we begin with, the high energy density of states scales as 
\be\label{cardy}
S \sim E^{\frac{d-1}{d}} \ .
\ee
The deformation shifts the energies, and changes the entropy accordingly. Denote by $E_0(\lambda,E)$ the initial energy of a state that has energy $E$ after the flow, which is easily calculated by inverting \eqref{cften}. Inputting into \eqref{cardy} gives the entropy of the flowed theory, 
\be\label{newent}
S(\lambda,E) \sim  E_0(\lambda,E)^{\frac{d-1}{d}} \sim \left(E +  E^2 L b\right)^{\frac{d-1}{d}}
\ee
with $L$ the system size and $b=-\frac{d}{d-1}\frac{\lambda}{L^d} >0$ a dimensionless parameter. For $EL b \ll 1$ the entropy reduces to the extensive scaling indicative of a local QFT, while for $EL b \gg 1$ the entropy becomes
\be
S \sim E^{\f{2(d-1)}{d}}.
\ee
Notice that this scaling is Hagedorn for $d=2$, as discussed in \cite{Giribet:2017imm,Giveon:2017nie, Giveon:2017myj}, and super-Hagedorn for $d>2$. Interestingly, this super-Hagedorn scaling matches the density of states of $(d-1)$-branes in the semiclassical approximation \cite{Duff:1987cs, Fubini:1972mf, Strumia:1975rd, Alvarez:1991qs}. The black holes in such a theory would have negative specific heat, like those in flat space. In fact, for $d=4$ the entropy scaling matches that of five-dimensional Schwarzschild black holes in flat space. It would be fascinating if the quantum theory defined by the irrelevant $T^2$ deformation considered here gave a new route to quantization of a theory of membranes.

\subsubsection*{The $\l > 0$ theory}
Equation \eqref{newent} does not apply to generic theories with $\l > 0$. This is because the CFT formula \eqref{cardy} is generically an asymptotic formula, while the $\l > 0$ deformation makes energies above an $E_{\text{max}}$ complex, as in \eqref{cften}. However, for holographic theories, or alternatively for modular invariant theories with a particular pattern of center symmetry breaking \cite{Shaghoulian:2016xbx}, this formula has an extended range of validity \cite{Shaghoulian:2015lcn}, holding down to energies $-(d-1)E_{\text{vac}}$. (For $d=2$ this extended range is equivalent to a sparse light spectrum \cite{Hartman:2014oaa}; for the connection to a sparse light spectrum in $d>2$ see \cite{Belin:2016yll, Shaghoulian:2016xbx}.) This means that the ensuing formulas can be applied to the $\l>0$ theory for states in the window $-(d-1)E_{\text{vac}}<E<E_{\text{max}}=(d-1)L^{d-1}/(4d\l)$.

An intriguing aspect of the deformation considered is that it preserves center symmetry for theories where it is present. It was argued in \cite{Shaghoulian:2016xbx} that the presence and pattern of spontaneous breaking of this symmetry is a robust way of reproducing aspects of semiclassical bulk physics when the boundary theory is placed on nontrivial topology. For example, the fact that the symmetry is unbroken means we can write correlation functions on quotient spacetimes (at leading order in $N$) in terms of a sum over images of the correlation function in the original spacetime; this important property is manifest from the bulk description, and in our dual EFT is kept intact by the preservation of center symmetry along the $T^2$ flow. 

\subsection{Speed of sound}
The speed of sound in these theories also shows interesting behavior. Fixing to flat space and using the pressure $p = \f{1}{(d-1)L^{d-2}}\f{dE}{dL}$ and the energy levels  \eqref{cften}, we find 
\be \label{SOS}
c_s = \sqrt{\frac{\partial p}{\partial \rho}} = \left(\frac{1}{d-1}\right)^{1/2}\left(\frac{1+(d-2)M\tilde{\l}}{1-2M\tilde{\l}}\right)^{1/2}\,,
\ee
where $\tilde{\l} = \frac{2d\l}{(d-1)L^d}$. In the $\tilde{\l} \to 0$ limit this reduces to the usual result, $c_s = 1/\sqrt{d-1}$. Moreover, the function is monotonic and diverges precisely at $M_{\text{max}}$ set by the square-root singularity in the energies. Hence for any positive $\l$ there exist finite-temperature states set by $M$ for which the speed of sound becomes arbitrarily large. This behaviour is identical to the two dimensional case \cite{McGough:2016lol}. The speed of sound in a theory with $\tilde{\l}<0$ needs to be interpreted with care, since the above formula is a thermodynamic formula. As seen in the previous subsection, the $\tilde{\l}<0$ theory has a super-Hagedorn density of states, so the canonical ensemble is ill-defined at any temperature.

In the bulk, the computation of the speed of sound in AdS with a Dirichlet wall at $r = r_c$ was done in \cite{Brattan2011}. They find
\be 
c_s^2 = \f{1}{d-1}\left(1+\f{d}{2\left(r_c^d/r_0^{d/(d-2)}-1\right)}\right).
\ee
Using \eqref{MassChargeBB} and \eqref{mu} to trade $r_0$ and $r_c$ for $M$ and $\l$, we see that this matches exactly with the field theory speed of sound found in \eqref{SOS}.

\section{Two-point functions}

So far, we have computed the spectrum and certain thermodynamic quantities of the deformed theory and found that they match with the dual bulk computation. To understand the role of the background terms, and demonstrate how the dictionary works more generally, we will also compute and compare 2-point correlation functions. In section \ref{scalarcorr} this was already done for scalar correlators. Here we will compute the flow of two-point functions of conserved $U(1)$ currents and stress tensors. The results will agree with the bulk calculation at finite cutoff. We will limit the discussion to vacuum two-point functions on flat space.  

\subsection{$U(1)$ current correlators}\label{u1}
Conserved $U(1)$ currents arise from gauge fields in the bulk. We have seen in section \ref{addingMatter} that such gauge fields give rise to two terms in the deformation, which are the analogues of $\partial J$ and $\cal O$ seen in \eqref{scalarflowp} for the scalar case. In particular, the flow of the effective action is
\be
\frac{\partial W[A]}{\partial \l} = \int d^dx \left(X_0 + \frac{\sigma_1}{\l^{2/d}}J^{i}J_{i} - \frac{\sigma_2}{\l^{\frac{2(d-2)}{d}}}F_{ij}F^{ij}\right),
\ee 
where $F_{ij} = \partial_{i} A_{j} - \partial_{j}A_{i}$ and $\sigma_i$ the dimensionless constants found in \eqref{gammas}. The operator $X_0$ is the deformation for gravity only. We now wish to compute the flow of the current two-point function by taking functional derivatives with respect to $A$,
\be
\partial_\l \braket{J^{l}(x)J^{m}(y)} = \frac{\d }{\d A_{l}(x)}\frac{\d}{\d A_{m}(y)}\Bigg\langle \int d^dy' \left(-\frac{\sigma_1}{\l^{2/d}}J^{i}J_{i} + \frac{\sigma_2}{\l^{2(d-2)/d}}F_{ij}F^{ij}\right)\Bigg\rangle.
\ee
Using 
\be
\braket{J^{l}(x)J^{m}(y)} = \frac{\d \braket{J^{m}(y)}}{\d A_{l}(x)}
\ee
and taking the large $N$ limit, the flow equation for the current correlator can be written as
\be\label{flowA}
\partial_\l \braket{J^{l}(x)J^{m}(y)} = 2\hspace{-1mm}\int \hspace{-1mm}d^d y' \left(\hspace{-1mm}-\frac{\sigma_1}{\l^{2/d}} \braket{J^{l}(x)J_{i}(y')} \braket{J^{i}(y')J^{m}(y)} + \frac{\sigma_2}{\l^{2(d-2)/d}}\Bigg\langle\frac{\d F_{ij}(y')}{\d A_{l}(x)}\frac{\d F^{ij}(y')}{\d A_{m}(y)}\Bigg\rangle\right).
\ee 
This flow equation simplifes in momentum space, where Lorentz invariance forces the two point function of $J^i$ to be of the form \cite{Bzowski2014}
\be
\braket{J^{l}(\mathbf{k})J^{m}(-\mathbf{k})} = C(\l,k)\pi^{lm},\quad \pi^{lm} = \delta^{lm} - \frac{k^{l}k^{m}}{k^2},
\ee
with $C$ a function of $\l$ and $k$ that completely fixes the two-point function. We have also stripped the delta function enforcing momentum conservation, just as in the scalar case. Plugging this in \eqref{flowA}, we find
\be\label{flowC}
\partial_\l C(\l,k) = -\frac{2\sigma_1}{\l^{2/d}} C(\l,k)^2 + \frac{4k^2\sigma_2}{\l^{2(d-2)/d}}.
\ee
Notice that this flow was also found in \cite{Heemskerk:2010hk}. This differential equation is supplemented with the CFT initial condition as $\l \rightarrow 0$, which, in position space, is just \eqref{JJcorrelator}. The solution is then
\be\label{solutionA} 
C(\l,k) = - \sqrt{\frac{2\sigma_2}{\sigma_1}} k \l^{\frac{3-d}{d}}\frac{K_{d/2-2}(2\sqrt{2\sigma_1\sigma_2}k \l^{1/d} d)}{K_{d/2-1}(2\sqrt{2\sigma_1\sigma_2}k \l^{1/d} d)}\,,
\ee
with $K$ the modified Bessel function of the second kind. When we insert the expressions for $\sigma_i$ to write this in terms of $r_c$, we find an exact match with the bulk computations done in \cite{MuckGauge}. Let us study the $d=3$ case in a bit more detail. Using the values of $\sigma_i$ given in \eqref{gammas}, we find that the correlator is given by
\be\label{d3Jflow}
\braket{J^{l}(\mathbf{k})J^{m}(-\mathbf{k})} = -\frac{C_J}{4\pi}k\pi^{lm}\,,
\ee
which is precisely the (Fourier transform of) the initial CFT value \eqref{JJcorrelator}. Thus for $d=3$ the correlator does not flow. As explained at the end of section \ref{scalarcorr}, this is due to the fact that the bulk theory is conformal in this case.

In even dimensions \eqref{solutionA} contains logarithms and to implement the initial condition as $\l \to 0$ it is convenient to analytically continue in $d$ and do the Fourier transform to position space, just as is done in \cite{MuckGauge}. The $\l \to 0$ limit is singular for $d>4$, but this simply reflects the fact that there are additional counterterms that we have neglected to include. Including them via our procedure will result in a finite answer.

\subsection{Stress-tensor correlators}
Let us now consider correlators of the stress tensor at finite $\l$ which we will show are dual to the propagator of gravitational perturbations at some constant $r = r_c$ surface in the bulk. We will start with the field theory computation and compare that with the bulk calculation afterwards.

Stress tensor correlators are computed by taking functional derivatives of the effective action $W=-\log Z$,
\be 
\langle T_{i_1 j_1}(x_1) \cdots T_{i_n j_n}(x_n) \rangle = \frac{2}{\sqrt{\g}}\frac{\d }{\d \g^{i_1j_1}(x_1)} \cdots \frac{2}{\sqrt{\g}}\frac{\d }{\d \g^{i_nj_n}(x_n)} (-W[\g,\l])\,.
\ee
Again, we will focus on the two-point function of $T_{ij}$ on $\mathbb{R}^d$ in the vacuum. Moreover, as we need the deformation for a general curved background to compute the correlators, we will only consider two, three and four boundary dimensions. As explained in section \ref{deriv} our derivation works generally, but becomes more tedious in $d > 4$. Our deformation is
\ba
\frac{\partial W}{\partial \lambda} =\int d^dx \sqrt{\gamma} \left[\left(T_{ij} + \frac{\alpha_d}{\lambda^{\frac{d-2}{d}}}  G_{ij}\right)^2 - \frac{1}{d-1}\left(T_i^i +\frac{\alpha_d}{\lambda^{\frac{d-2}{d}}} G_i^i\right)^2 \right],
\ea  
where $\alpha_d = (d^{1-2/d} (d-2)(\pi G)^{2/d} 2^{1+4/d})^{-1} $. To compute the flow of the stress-tensor two-point function, we proceed analogously as for the gauge field. We go to momentum space, where stress tensor two-point functions in the vacuum can be written in terms of following two tensor structures (again omitting the overall delta function which enforces momentum conservation)
\bea\label{eq:struc}
&&\langle T_{ij}(\mathbf{k}) T_{lm}(-\mathbf{k}) \rangle_\lambda = A(k,\lambda) \Pi_{ijlm} +B(k,\lambda) \pi_{ij} \pi_{lm},  \\
&& \pi_{ij} = \d_{ij} - \frac{k_i k_j}{k^2}, \qquad  \Pi_{ijlm}  = \frac{1}{2} (\pi_{il} \pi_{jm} +\pi_{im} \pi_{jl})  - \frac{1}{d-1} \pi_{ij} \pi_{lm} .
\eea
Note that in $d=2$ the first structure $\Pi_{ijlm}$ vanishes identically. Taking derivatives with respect to the metric and decomposing the expression in terms of $\Pi_{ijlm}$ and $\pi_{ij}\pi_{lm}$ we find
\be\label{eq:ab}
\pa_{\lambda} A(k,\l) =  -2  \left(A(k,\l) -  \alpha_d k^2 \l^{-(d-2)/d} \right)^2, \qquad \pa_\lambda B(k,\l) = 0\,,
\ee 
where in deriving the above equations we kept leading terms in $1/N$ and used $\langle T_{ij}\rangle =0$. The identity
\bea
\f{\d G_{ij}}{\d g^{lm}} &&= \f{1}{4}\left(k^2 \d_{lm}\pi_{ij}-k^2\d_{jm}\pi_{il}+\d_{im}k_l k_j - \d_{ij}k_l k_m\right)+l\leftrightarrow m\,, \\
&& = \frac{d-2}{2(d-1)} k^2 \pi_{ij} \pi_{lm} -\frac{1}{2}k^2 \Pi_{ij lm}
\eea
which leads to 
\be
\f{\d G_{pq}}{\d g^{ij}}\f{\d G^{pq}}{\d g^{lm}}-\f{1}{d-1}\f{\d G^p_p}{\d g^{ij}}\f{\d G^q_q}{\d g^{lm}} =\f{k^4}{4}\Pi_{ijlm}\,,
\ee
is useful in deriving the above. The Ricci scalar term present in the trace relation for $d=2$ is topological once integrated and does not contribute to the correlation functions.

The constancy of $B$ under the flow of the deformation has the following consequence. Upon taking the trace of \eqref{eq:struc} we find that
\be
\langle T_{i}^{i}(\mathbf{k}) T_{m}^m(\mathbf{-k})\rangle_\lambda = B(k, \lambda) (d-1)^2.
\ee
In $d=2$, this is proportional to central charge, therefore in a $T\bar{T}$ deformation of holographic CFTs, \eqref{eq:ab} immediately implies that the central charge does not flow. This is consistent with both the field theory result \cite{Aharony:2018vux} and the bulk gravity computation \cite{Kraus:2018xrn}. Also, in any odd dimensional CFT, there is no anomaly and hence $B(k,\l)=0$. In even dimensions, there is a trace anomaly, but by expanding the trace anomalies around the Minkowski spacetime we find $B(k, \l) =0$ in $d \ge 4$.

The solution for $A(k,\lambda)$ is given by
\be\label{eq:Ak}
A(k,\lambda) = -\frac{\tilde{k} \l^{1/d-1} }{2d} \frac{K_{1-d/2}(\tilde{k} \lambda^{1/d})}{K_{d/2}(\tilde{k} \lambda^{1/d})} +\alpha_d k^2 \l^{-(d-2)/d}\,,
\ee
where $K$ is the modified Bessel function of the second kind and $\tilde{k} = d \sqrt{\frac{2(d-2) \alpha_d}{d}} k =  \left(\frac{d}{4\pi G}\right)^{1/d} k$. Note that due to the second term above, this solution has a smooth limit as $\l \to 0$ and matches exactly onto the CFT answer for $ d \leq 4$.  In $d=3$ the form of the two point function is simple and given by

\be
\left. A(k,\lambda) \right|_{d=3}  = \frac{\sqrt{6} \alpha_3^{3/2} k^3  }{1+ k \sqrt{6\alpha_3} \l^{1/3}}\,.
\ee
As in the case of gauge fields, even dimensions have logarithms in the small $\l$ limit. For $d>4$ one needs to add more counterterms to find a smooth limit as $\l \to 0$. Our result is in agreement with the known bulk result for the two-point function of the stress tensor \cite{Mueck:1998ug}. However, note that in \cite{Mueck:1998ug} the Einstein tensor counterterm is absent and computing the on-shell action gives only the first term in \eqref{eq:Ak}. In that approach, the correct correlator is found by dropping local terms arising from the Bessel functions, whereas in our approach local terms cancel with counterterms and the correlator has the correct power law behaviour when $\l \to 0$.

\section{Discussion}

We have studied effective field theories defined by the flow \eqref{introflow}. The calculability of quantities like the deformed energy spectrum and correlation functions came from our assumption of large-$N$ factorization. The operator defining the flow was extracted by considering bulk AdS physics; in this context we provided evidence that the dimensionful parameter $\l$ is related to a sharp radial cutoff in the bulk. 

An important challenge facing development of this approach is $1/N$ corrections. These are essential to gain a better handle on quantum gravity in finite patches of spacetime. Our deforming operator was selected by a bulk classical analysis, which can be modified by quantum effects.

Another interesting direction to pursue is the case of $d=1$. The techniques we use are general and can be applied to e.g. Jackiw-Teitelboim gravity in two dimensions. In the limit where the cutoff is taken to be close to the boundary, the deformed theory should correspond to the Schwarzian theory \cite{Jensen:2016pah,Maldacena:2016upp,Engelsoy:2016xyb}.

As mentioned in the introduction, the gravity theory with a Dirichlet cutoff is rather strange, and the dual EFT is correspondingly strange. One possibility is that the theory makes sense only as an `intermediate step' in a bigger calculation. For example, the full AdS/CFT duality can, in principle, be cut at some arbitrary surface $r = r_c$, then recovered by integrating over all fields at the cutoff, including the metric.  (See for example in \cite{Heemskerk:2010hk}.) In this calculation, the bulk partition function with finite cutoff appears in the first step, but the dual EFT is then coupled to gravity and to another theory in the UV. This is similar to the role of the wavefunction in the dS/CFT correspondence as formulated in \cite{Maldacena:2002vr} --- there, the wavefunction of the universe is calculated with a Dirichlet boundary condition at fixed time, but physical observables are obtained only after integrating over boundary conditions. In the AdS case, this suggests that although our EFT may not make sense in the UV as a quantum field theory, it should be possible to UV-complete the theory when coupled to gravity. (This suggests the existence of an anti-swampland: a class of effective field theories that cannot be UV-completed \textit{unless} coupled to gravity!)

\section*{Acknowledgments}
It is a pleasure to thank Dionysios Anninos, Jan de Boer, Sergei Dubovsky, Zohar Komargodski, Raghu Mahajan, Mark Mezei, Herman Verlinde, and Sasha Zamolodchikov for useful conversations. TH and AT are supported by Simons Foundation grant 488643. JK is supported by the Delta ITP consortium, a program of the Netherlands Organisation for Scientific Research (NWO) that is funded by the Dutch Ministry of Education, Culture and Science (OCW). ES is supported in part by NSF grant no. PHY-1316748 and
Simons Foundation grant 488643. TH and ES also acknowledge support from the Aspen Center for Physics, which is supported by NSF grant PHY-1607611, during the program ``Information in Quantum Field Theory" in Summer 2017.

\bigskip\appendix
\section{Conventions}
Our conventions, in Euclidean signature, are as follows. The path integral is over $e^{-S}$. The bulk action is
\be
S = - \frac{1}{16\pi G}\int_M\sqrt{g}(R-2\Lambda) - \frac{1}{8\pi G}\int_{\partial M}\sqrt{g^0}K + S_{ct}  + S_{matter}\ .
\ee
The gravitational counterterm is 
\be
S_{ct} = \frac{1}{8\pi G}\int_{\partial M}\sqrt{g^0}(d-1 + {\cal L}_{curv}) ,
\ee
where in $d=3,4$ the curvature counterterm is ${\cal L}_{curv} = \frac{1}{2(d-2)}R[g^0]$ \cite{Balasubramanian1999}, and higher dimensional curvature counterterms can be found in \cite{Emparan:1999pm}. The extrinsic curvature is $K_{\mu\nu} =  2\nabla_{(\mu}n_{\nu)}$ with $n$ the outward-pointing normal. The Brown-York stress tensor is defined by
\be
\delta S = \frac{1}{2}\int_{\p M}\sqrt{g^0}\, \tbar^{\mu\nu}\delta g_{\mu\nu}^0 \ ,
\ee
and the convention for the stress tensor in the boundary theory is similar.

Our sign conventions for the Euclidean generating functional in the EFT are summarized by
\be
Z[J, h_{ij}, A_i] = \left\langle e^{\int d^dx \sqrt{\gamma}(J {\cal O}  +  A_iJ^i- \half h_{ij}T^{ij} ) }\right\rangle \ .
\ee
This choice, together with our sign choices in the bulk action, produces positive boundary two-point functions in position space, for example  $\langle {\cal O}(x) {\cal O}(y)\rangle = \left(\frac{1}{\sqrt{\gamma}} \frac{\delta}{\delta J}\right)^2 \log Z = |x-y|^{-2\Delta}$. This is subtle due to divergences in the Fourier transform from momentum to position space; for $\Delta > d/2$, the calculation is done in momentum space, and the Fourier transform to position space is done by analytic continuation in $\Delta$, or by putting a hard cutoff $|k| < \Lambda$ and adding local counterterms to eliminate divergences.

\small
\bibliographystyle{ourbst}
\bibliography{TTbarBib}

\end{document}